\documentclass[twocolumn,showpacs,prb,amsfonts,amsmath,amssymb,floatfix,groupedaddress]{revtex4-2}
\usepackage{color}
\usepackage{hhline}
\usepackage{mathrsfs}
\usepackage{graphicx}
\usepackage{dcolumn}
\usepackage{bm}
\usepackage{multirow}
\usepackage{booktabs}
\usepackage{afterpage}
\usepackage{amsmath}

\begin{document}

\title{Revisiting the homogeneous electron gas in pursuit of the properly normed {\it ab initio} Eliashberg theory}

\author{Ryosuke Akashi$^{1}$}
\thanks{ryosuke.akashi@phys.s.u-tokyo.ac.jp}
\affiliation{$^1$Department of Physics, The University of Tokyo, Hongo, Bunkyo-ku, Tokyo 113-0033, Japan}

\date{\today}
\begin{abstract}
We address an issue of how to accurately include the self energy effect of the screened electron-electron Coulomb interaction in the phonon-mediated superconductors from first principles. In the Eliashberg theory for superconductors, self energy is usually decomposed using the $2\times 2$ Pauli matrices in the electron-hole space. We examine how the diagonal ($\sigma_{0}$ and $\sigma_{3}$) components resulting in the quasiparticle correction to the normal state, $Z$ and $\chi$ terms, behave in the homogeneous electron gas in order to establish a norm of treating those components in real metallic systems. Within the $G_{0}W_{0}$ approximation, we point out that these components are non-analytic near the Fermi surface but their directional derivatives and resulting corrections to the quasiparticle velocity are nevertheless well defined. Combined calculations using the $G_{0}W_{0}$ approximation and Eliashberg equations show us that the effective mass and pairing strength strikingly depend on both $Z$ and $\chi$, in a different manner. The calculations without the numerically demanding $\chi$ term is thus shown to be incapable of describing the homogeneous electron gas limit. This result poses a challenge to accurate first-principles Eliashberg theory.
\end{abstract}

\maketitle

\section{Introduction}
The Eliashberg theory is a prevalent framework to theoretically examine low-temperature electronic properties of superconductors~\cite{Eliashberg1960,Scalapino-bookchap,Schrieffer-book}. Its surprising capability has continuously been demonstrated through successful explanations of various observable quantities of superconductors such as excitation spectra and thermodynamic values~\cite{Carbotte-RMP1990}.
With continuous progress in the first-principles calculation methods, this theory has gained remarkable accuracy to describe the physics of the phonon-mediated superconductors. One of its recent triumphs is consistent explanation of the hydride superconductors~\cite{Flores-Livas2020}. The building blocks of the theory are microscopic electronic and phononic properties of the target system. Calculating them for crystal structures predicted by efficient searching algorithms enables us to reproduce the experimentally observed superconducting properties accurately~\cite{Oganov-book}.

Now a trend is observed that the other factor of the Eliashberg theory, the electron-electron Coulomb interaction, also be analyzed from the first principles. In the phonon-mediated superconductors, previously, only one aspect of its effect has been focused that it counteracts the phonon-mediated pairing attraction and suppresses the superconducting transition. It has been practically treated as a single semiempirical parameter $\mu^{\ast}$ (Ref.\onlinecite{Morel-Anderson}). Nonempirical determination of its absolute impact on $T_{\rm c}$ with practically reduced computational cost has become feasible with the advent of the superconducting density functional theory~\cite{OGK, SCDFTI, SCDFTII}. Solution of the Eliashberg equation free from the parameter $\mu^{\ast}$ (Refs.~\onlinecite{Sanna2018,Wang-Eliashberg}) is also getting into practice in recent few years. 

Moreover, recent first-principles studies have shed light on the qualitatively different aspects of the Coulomb interaction. Intriguing examples are the correlation-enhanced electron-phonon coupling~\cite{Janssen2010,Yin2013, Calandra2015,Louie2019}. and spin-fluctuation effects~\cite{Essenberger2014, Essenberger2016,Kawamura2020,Tsutsumi2020}. Another one, which is indeed the target of this paper, is the plasmonic effect. The plasmon-mediated pairing in the electron gas has first been proposed by Takada~\cite{Takada1978} decades ago. Its remarkable ability to co-operate with the phonon-mediated pairing has recently been featured by Akashi and Arita~\cite{Akashi2013PRL,Akashi2014JPSJ} with an extension of the superconducting density functional theory. On the other hand, Davydov and coworkers~\cite{Davydov2020} have argued that the pairing effect thus included may be so strong as to yield artificially large Tc, which would be mitigated by the electronic mass renormalization effect by the plasmons.

We are thus witnessing rapid advances in the first-principles simulation of the electron-electron Coulomb effect on superconductors, which must have immediate impact in modern superconducting studies. We thereby motivate ourself to delve into the problem of incorporating the plasmonic effect in the first-principles Eliashberg theory. The plasmonic effect is well represented by the electronic self energy due to the dynamically screened Coulomb interaction within the random-phase approximation (RPA)~\cite{Pines-RPA}. Its study in the normal state, especially of the homogeneous electron gas, has a long history. However, its full treatment in the Eliashberg calculations has yet been unprecedented as described below. Different levels of approximations have been practiced in these two contexts, the relation of which has been unclear. Our aim is to fill this gap by reinterpreting the former well-established result in the context of the latter as a firm basis for future calculations in real systems.

In this paper, we reanalyze the RPA self energy in the homogeneous electron gas in the language of the Eliashberg theory. The numerical difficulties in solving the Eliashberg equations with the full RPA screened Coulomb interaction and the current standard approximations are first summarized. The drawback of the approximation is next examined in the homogeneous electron gas, where we can refer to some analytical results. Insights into the real metallic systems are thereby derived and preferable ways to manage the drawback is discussed.

Specifically, we calculate the normal-state self energy with the non-self-consistent G$_{0}$W$_{0}$ approximation~\cite{Hedin1965}. The Eliashberg gap equation is solved using the self energy, with the phonon effect treated with the Einstein model. Impact of low-energy singularities is carefully examined and confirmed to be irrelevant. The main findings on the effects of the screened Coulomb interaction are as follows. The mass renormalization is negligibly small due to cancellation of large components having opposite signs. On the other hand, the pairing strength is significantly suppressed in total. Deriving an analytical formula, we point out that the renormalized quasiparticle weight, not the effective mass, is responsible for the latter. With the disparity of mechanisms of these two, the current standard approximations used in the first-principles Eliashberg calculations are found to yield improper behaviors in the homogeneous limit for either the mass and pairing renormalizations.

\section{Theory}
Although we finally perform analyses only for the homogeneous electron gas, we start from the general inhomogeneous electron-phonon Hamiltonian for phonon-mediated superconductors
\begin{eqnarray}
\hat{H}=
\hat{K}+\hat{V}_{\rm ion}+\hat{H}_{\rm ph}+\hat{V}_{\rm el-ph}+\hat{U}_{\rm el-el}
-\mu_{0}\hat{N}
,
\end{eqnarray}
where $\hat{K}$ denotes the electron kinetic term and $\hat{V}_{\rm ion}$ is the ionic potential. $\hat{H}_{\rm ph}$ represents the phononic Hamiltonian. $\hat{V}_{\rm el-ph}$ and $\hat{U}_{\rm el-el}$ denote the electron-phonon coupling and electron-electron Coulomb interaction, respectively. $\mu_{0}$ and $\hat{N}$ are the chemical potential and total number operator, respectively. In the Eliashberg theory, the electronic thermal Green's function is expressed conveniently with the Nambu-Gorkov formalism~\cite{Nambu1960,Gorkov1958}, a 2$\times$2 matrix form as
\begin{eqnarray}
{\bm G}(1,2) = -
\left\langle
T \left[
\left(
\begin{array}{c}
\psi(1) \\
\psi^{\dagger}(1)
\end{array}
\right)
\times
\left(
\psi^{\dagger}(2) \ 
\psi(2)
\right)
\right]
\right\rangle
.
\label{eq:Green}
\end{eqnarray}
$\psi$ denotes the electronic annihilation operator in the imaginary-time Heisenberg representation. Arguments ``1", ``2", $\cdots$ abbreviate the space-imaginary time points $({\bm x}_{1}, \tau_{1}), ({\bm x}_{2}, \tau_{2}), \cdots$. Operator $T$ denotes the fermionic time ordering and the average $\langle\  \rangle$ is taken for the interacting equilibrium ensemble. 

The Green's function is related to a non-interacting reference Green's function ${\bm G}_{0}$ by the Dyson equation
\begin{eqnarray}
&&{\bm G}({\bm r}_{1}, {\bm r}_{2}, \omega_{j})
=
{\bm G}_{0}({\bm r}_{1}, {\bm r}_{2}, \omega_{j})
\nonumber \\
&&
+
\int d{\bm r}_{3}d{\bm r}_{4}
{\bm G}_{0}({\bm r}_{1}, {\bm r}_{3}, \omega_{j})
{\bm \Sigma}({\bm r}_{3}, {\bm r}_{4}, \omega_{j})
{\bm G}({\bm r}_{4}, {\bm r}_{2}, \omega_{j}).
\nonumber \\
&&
\end{eqnarray}
Here the Fourier transformation to Matsubara frequency $\omega_{j}$ has been executed. The non-interacting Green's function ${\bm G}_{0}$ is also defined by Eq.~(\ref{eq:Green}) but with the non-interacting Hamiltonian
\begin{eqnarray}
\hat{H}_{0}=
\hat{K}+\hat{V}_{\rm ion}+\hat{V}_{\rm eff}
-\mu_{0}\hat{N}
,
\end{eqnarray}
The operator $\hat{V}_{\rm eff}$ is the effective one-body potential which presumably includes the dominant part of the electron-electron interaction effects, which may in principle be either local or nonlocal. In the standard first-principles calculations $\hat{V}_{\rm eff}$ is taken to be the exchange-correlation potential~\cite{Kohn-Sham}.

The self energy is decomposed into the Pauli matrices as
\begin{eqnarray}
&&{\bm \Sigma}
(i\omega_{j})
=i\omega_{j}(1-Z(i\omega_{j}))\sigma_{0}
\nonumber \\
&&\hspace{40pt}
+\phi(i\omega_{j})\sigma_{1}+(\chi(i\omega_{j})-V_{\rm eff})\sigma_{3}
.
\end{eqnarray}
The $\sigma_{2}$ component is suppressed to zero by the gauge transformation. The nondiagonal ($\sigma_{1}$) component becomes nonzero in the superconducting phase, which can be regarded as the superconducting order parameter. The diagonal ($\sigma_{0}$ and $\sigma_{3}$) ones represent the renormalization of the quasiparticle spectrum which is relevant also in the normal state. With the term $-V_{\rm eff}\sigma_{3}$, the doubly counted interaction effect is subtracted.  The specific form of ${\bm \Sigma}$ depends on the interactions and approximations used. Note that from the definition of the $2\times 2$ Green's function, only two components of the self energy matrix are independent
\begin{eqnarray}
{\bm \Sigma}
=
\left(
\begin{array}{cc}
\Sigma  & F \\
F^{\ast} & -\Sigma^{\ast}
\end{array}
\right)
\end{eqnarray}
where $\ast$ denotes the time reversal counterpart of the quantities. In the time-reversal symmetric systems, the diagonal components are therefore related to the self energy in the single component notation $\Sigma$ as~\cite{Allen-Mitrovic,Marsiglio-review}
\begin{eqnarray}
i\omega_{j}(1-Z(i\omega_{j}))
&=&\frac{1}{2}
\left[
\Sigma(i\omega_{j})
-
\Sigma(-i\omega_{j})
\right]
\label{eq:sigma-Z-T}
\\
\chi(i\omega_{j})
-V_{\rm eff}
&=&
\frac{1}{2}
\left[
\Sigma(i\omega_{j})
+
\Sigma(-i\omega_{j})
\right]
\label{eq:sigma-chi-T}
\end{eqnarray}

In the Eliashberg theory, the following term is considered as the self energy~\cite{Schrieffer-book}
\begin{eqnarray}
{\bm \Sigma}({\bm r}_{1}, {\bm r}_{2},\omega)
&=&
i
\int d\omega'
\sigma_{3}
{\bm G}({\bm r}_{1}, {\bm r}_{2},\omega-\omega')
\sigma_{3}
W({\bm r}_{1}, {\bm r}_{2},\omega')
\nonumber \\
&&-V_{\rm eff}({\bm r}_{1}, {\bm r}_{2})\sigma_{3}
\end{eqnarray}
with $W$ being the total effective interaction $W=W^{\rm ph}+W^{\rm el}$; $W^{\rm ph}$ originates from the dressed phonon propagator and screened electron-phonon vertices and $W^{\rm el}$ denotes the RPA screened Coulomb interaction, respectively. The neglect of higher-order terms with respect to the phonon propagator is justified by the Migdal theorem~\cite{Migdal1958}. On the other hand, there is no general theorem which assures those with respect to the screened Coulomb interaction are small, except for the trivial dense limit where electrons are asymptotically free. 

Assuming that the self energy is diagonal with respect to the basis, the Dyson equation is recast to the following set of self-consistent equations (Eliashberg equations), 
\begin{eqnarray}
Z_{i}(i\omega_{j})
&=&
1
-
T
\sum_{i'j'}
\frac{\omega_{j'}}{\omega_{j}}\frac{Z_{i'}(i\omega_{j'})}{\Theta_{i'}(i\omega_{j'})}
W_{ii'}(i(\omega_{j}-\omega_{j'}))
,
\label{eq:Eliashberg-Z}
\\
\chi_{i}(i\omega_{j})
&=&
T
\sum_{i'j'}
\frac{\xi_{i'}+\chi_{i'}(i\omega_{j'})}{\Theta_{i'}(i\omega_{j'})}
W_{ii'}(i(\omega_{j}-\omega_{j'}))
,
\label{eq:Eliashberg-chi}
\\
\phi_{i}(i\omega_{j})
&=&
-
T
\sum_{i'j'}
\frac{\phi_{i'}(i\omega_{j'})}{\Theta_{i'}(i\omega_{j'})}
W_{ii'}(i(\omega_{j}-\omega_{j'}))
,
\label{eq:Eliashberg-phi}
\\
\Theta_{i}(i\omega_{j})
&=&
\left[
\omega_{j}
Z_{i}(i\omega_{j})
\right]^2
\!+\!
\left[
\xi_{i}
+
\chi_{i}(i\omega_{j})
\right]^2
\!+\!
\left[
\phi_{i}(i\omega_{j})
\right]^2
.
\nonumber \\
\end{eqnarray}
Indices $i$ ($i'$) and $j$ ($j'$) denote the basis state and Matsubara frequency, respectively. $\xi_{i}$ is the energy eigenvalue of the state $i$ measured from the chemical potential $\mu_{0}$. $\mu_{0}$ should be tuned so that the number conservation condition is satisfied~\cite{Marsiglio-review}, though this step is mostly not executed practically. The matrix element of the interaction $W_{ii'}(\nu_{j})$ for the Bosonic Matsubara frequency $\nu_{j}$ is defined between the Cooper-pair states 
\begin{eqnarray}
&&W_{ii'}(\nu_{j})
\nonumber \\
&&=\int d{\bm r}_{1}d{\bm r}_{2}
\psi^{\ast}_{i}({\bm r}_{1})\psi^{\ast}_{i*}({\bm r}_{2})
W({\bm r}_{1}, {\bm r}_{2},\nu_{j})
\psi_{i'{\ast}}({\bm r}_{2})\psi_{i'}({\bm r}_{1})
.
\nonumber \\
\end{eqnarray}
Here the wavefunction $\psi_{i}$ diagonalizes the non-interacting Hamiltonian $\hat{H}_{0}$ and $i*$ denotes the state which is the time-reversal counterpart of $i$.

Usually the approximation $W\simeq W^{\rm ph}$ is applied to Eqs.(\ref{eq:Eliashberg-Z}) and (\ref{eq:Eliashberg-chi}). This treatment is based on the consideration that the effect of $W^{\rm el}$ on $Z$ and $\chi$ is mainly change of the one-particle spectrum from the bare to dressed one, which is presumably well accounted for if the basis set (or ${\bm G}_{0}$ in other words) is appropriately chosen~\cite{Schrieffer-book}. Note that our motivation is to revisit and ponder on the validity of this treatment.

The numerically tedious problem is condensed in $\chi$. Because of the factor of order $O(1/\xi_{j'})$, $[\xi_{i'}+\chi_{i'}(i\omega_{j'})]/\Theta_{j'}(i\omega_{j'})$, Eq.~(\ref{eq:Eliashberg-chi}) diverge unless the basis dependence of $W$ is explicitly treated. In most practices so far $\chi$ is ignored. The effect of $\chi$ is indeed minor as long as we consider only $W^{\rm ph}$, which decays by the phononic frequency scale, since it becomes zero when the electronic density of states is approximately constant around the chemical potential~\cite{Schrieffer-book}. When we consider $W^{\rm el}$ explicitly, on the other hand, this is obviously not the case. However, little is known about the $\chi$ part with the explicit treatment of $W^{\rm el}$ since its numerically case studies are lacking. 

Apart from the Eliashberg theory, the normal-state self energy of electron gas with $W^{\rm el}$ has been extensively studied. Gell-Mann~\cite{Gell-Mann1957-2} and Quinn and Ferrell~\cite{Quinn1958} have derived the correction to the effective mass in the homogeneous electron gas at zero temperature with the RPA to $W^{\rm el}$. Hedin has established a formal theory treating $W^{\rm el}$ as the perturbation~\cite{Hedin1965}. His proposal of the $GW$ approximation, where the vertex part is treated as bare one, has been implemented and widely applied to metals and semiconductors~\cite{Hybertsen-Louie1986,Godby1988,Aryasetiawan-review,Reining-review}. In this paper we aim to understand the characteristics of the $\chi$ term with the help of the knowledge of those accumulated studies. 

We study $Z$ and $\chi$ in the homogeneous electron gas by setting $\hat{V}_{\rm ion}$ and $\hat{V}_{\rm eff}$ to zero. The atomic unit is adopted: the electron charge and mass, speed of light, and Planck constant divided by $2\pi$ are unity. The basis states are taken to be the plane waves $\{e^{i{\bm k}\cdot {\bm r}}\}$. The system is characterized by the single parameter $r_{\rm s}=\sqrt[3]{3/(4\pi n)}$ with $n$ being the electron number density, or equivalently, the Fermi wavenumber $k_{\rm F}=1/(\alpha r_{\rm s})$ with $\alpha=(\frac{4}{9\pi})^{1/3}$. 

There are various simplified variants of the $GW$ approximation in practice~\cite{Aryasetiawan-review}. The full self consistency is accomplished by repeating the following cycle; (i) the calculation of $W=iGG$, (ii) $\Sigma=iGW$ and (iii) $G^{-1}=G^{-1}_{0}-\Sigma$. Executing this cycle only once starting from $G\equiv G_{0}$ is called $G_{0}W_{0}$ approximation. In solving the Eliashberg equations self-consistently, the update of $W$ is skipped; this corresponds to, say, the $GW_{0}$ approximation~\cite{Shishkin-Kresse-GW0}. In this work, we study the self energy effects with the $G_{0}W_{0}$ approximation. It is asymptotically exact in the dense limit, as well as analytic formula in the literature is available. Also, this approximation corresponds to the first cycle of the $GW$ and $GW_{0}$ approximations. Actually the accuracy of the self-consistent results has itself been a complicated problem~\cite{vonBarth-Holm1996,Shirley-GW,Aryasetiawan-review, Marom-Rubio-GW,Takada-GISC,Kutepov-GWG}; we do not address this point in this paper.

In Sec.~\ref{sec:mass} we examine the mass renormalization effects of $Z$ and $\chi$ in the normal state by ignoring $W^{\rm ph}$. We thoroughly reproduce the historically well received fact that the mass renormalization in the dense normal uniform electron gas is weak~\cite{Hedin-Lundqvist, Grimvall, Giuliani-Vignale} and reinterpret that in the Eliashberg context. The effects of $Z$ and $\chi$ on the superconducting instability is evaluated in Sec.~\ref{sec:single-Eliashberg} with a non-self consistent execution of the Eliashberg equations using a model function for $W^{\rm ph}$.

\subsection{Mass renormalization}
\label{sec:mass}
We move to the zero-temperature formalism. The resulting self-consistent equations for $Z$ and $\chi$ are identical to that of the single-component self enegy $\Sigma=iGW^{\rm el}$: the object of the $GW$ theory. Our approach is to analyze the latter and relate it to $Z$ and $\chi$, recalling the zero-temperature counterpart of the relations Eqs.~(\ref{eq:sigma-Z-T}) and (\ref{eq:sigma-chi-T})
\begin{eqnarray}
{\rm Re}[
\omega(1-Z(\omega))
]
&=&\frac{1}{2}
{\rm Re}
\left[
\Sigma(\omega)
-
\Sigma(-\omega)
\right]
\label{eq:sigma-Z}
\\
{\rm Re}[
\chi(\omega)
]
-V_{\rm eff}
&=&
\frac{1}{2}
{\rm Re}
\left[
\Sigma(\omega)
+
\Sigma(-\omega)
\right]
\label{eq:sigma-chi}
\end{eqnarray}

The $G_{0}W_{0}$ self energy in the uniform electron gas is given by
\begin{eqnarray}
&&\Sigma^{G_{0}W_{0}}(k, \omega)
=i\int \frac{d{\bm k}'d\omega'}{(2\pi)^4}G_{0}({\bm k}+{\bm k}', \omega+\omega')
W^{\rm el}({\bm k}', \omega')
\nonumber \\
\label{eq:sigma-G0W0}
\end{eqnarray}
with $W^{\rm el}({\bm k},\omega)=\frac{4\pi}{k'^2}\frac{1}{\varepsilon^{\rm RPA}({\bm k}', \omega')}$ and $\varepsilon^{\rm RPA}$ being the RPA (Lindhard) dielectric function~\cite{Lindhard1954}. The quasiparticle spectrum $E({\bm k})\equiv E(k)$ near the Fermi level is given by the pole of $G$ of $\omega$
\begin{eqnarray}
E(k)
=
E_{0}(k)
+
{\rm Re}\Sigma^{G_{0}W_{0}}(k, E(k))
\end{eqnarray}
where $E_{0}(k)=k^2/2-\mu_{0}$. Expanding $\Sigma(k, E(k))$ around $E_{0}(k)$, we get~\cite{Hedin1965}
\begin{eqnarray}
E(k)
\simeq
E_{0}(k)
+
\frac{{\rm Re}\Sigma(k, E_{0}(k))}{1-\left.\frac{\partial {\rm Re}\Sigma(k, \omega))}{\partial \omega}\right|_{\omega=E_{0}(k)}}
\label{eq:E-G0W0}
\end{eqnarray}
Up to the first order with respect to $\Sigma$, it reduces to the RPA formula~\cite{Quinn1958}
\begin{eqnarray}
E(k)
\simeq
E_{0}(k)
+
{\rm Re}\Sigma(k, E_{0}(k))
\label{eq:E-RPA}
\end{eqnarray}

In this subsection our interest is in the Fermi velocity $v_{\rm F}=dE(k)/dk|_{k=k_{\rm F}}$, or equivalently, effective mass $m^{\ast}\equiv 1/[d^{2}E(k)/dk^{2}|_{k=k_{\rm F}}]$. This is because the mass renormalization may have tremendous impact on the superconducting transition temperature. The McMillan-Allen-Dynes formula~\cite{McMillan, Allen-Dynes}, which is an approximate solution of the Eliashberg equations, 
includes $m^{\ast}$ in a form
\begin{eqnarray}
T_{\rm c}
\sim \Omega {\rm exp}
\left[
-\frac{m^{\ast}}{\lambda}
\right]
,
\label{eq:MAD}
\end{eqnarray}
where $\Omega$ and $\lambda$ respectively denote the typical phonon frequency and total effective coupling strength under the presence of $W^{\rm ph}$ and $W^{\rm el}$. Although the original equation only concerns the mass renormalization due to $W^{\rm ph}$, this formula suggests that the effect of $W^{\rm el}$ on $T_{\rm c}$ via the change of $m^{\ast}$ may be substantial. Precisely speaking, the appearance of the mass renormalization {\it factor} in the $T_{\rm c}$ formula does not mean that the mass renormalization {\it phenomenon} causes the $T_{\rm c}$ change: We revisit this point in the next subsection.

The Fermi velocity $v_{\rm F}=dE(k)/dk|_{k=k_{\rm F}}$ is determined by the two kinds of the $k$ dependence of $\Sigma$: on the first argument $k$ and second argument through $E_{0}(k)$. We show that those dependencies are related to $Z$ and $\chi$, respectively. 

The expansion form of $\Sigma^{G_{0} W_{0}}$ near $(k=k_{\rm F}, \omega=0)$ reads 
\begin{widetext}
\begin{eqnarray}
\Sigma^{G_{0} W_{0}}(k, \omega)
&\simeq&
\Sigma(k_{\rm F}, 0)
+
(k-k_{\rm F})
\left.
\frac{\partial \Sigma^{G_{0} W_{0}}(k, 0)}{\partial k}
\right|_{k=k_{\rm F}}
+
\omega
\left.
\frac{\partial \Sigma^{G_{0} W_{0}}(k_{\rm F}, \omega)}{\partial \omega}
\right|_{\omega=0}
\nonumber \\
&&
+
\frac{1}{2\pi}
\left(
\frac{\omega_{\rm p}}{\omega_{\rm p}-(\omega-\frac{k^2-k^2_{\rm F}}{2})}
+
\frac{\omega_{\rm p}}{\omega_{\rm p}+(\omega-\frac{k^2-k^2_{\rm F}}{2})}
-2
\right)(k-k_{\rm F}){\rm ln}|k-k_{\rm F}|
+\cdots
\label{eq:sigma-expand}
\end{eqnarray}
\end{widetext}
with $\omega_{\rm p}=\sqrt{3/r_{\rm s}^3}$ being the plasma frequency. Because of the fourth term, the usual Taylor expansion is not well defined, which has, to the author's knowledge, not been clearly appreciated in the literature. The function $\Sigma^{G_{0} W_{0}}(k, \omega)$ is not continuously differentiable around the reference point $(k=k_{\rm F}, \omega=0)$, nor interchangeable are the partial derivatives $\partial/\partial k$ and $\partial/\partial \omega$. The non-analytic term (see Appendix~\ref{app:singular} for derivation) originates from the long-range character of the screened Coulomb interaction with nonzero frequency. The $k$-derivative of this term diverges on the line $k=k_{\rm F}$, $\omega\neq 0$ whereas the $\omega$-derivative is finite. Note that this term gives smooth contributions of order $O((k-k_{\rm F})^2{\rm ln}|k-k_{\rm F}|)$ along the line $\omega=b(k-k_{\rm F})$ with an arbitrary parameter $b$ and of $O(\omega^2)$ along the line $k=k_{\rm F}$, both being smooth. Thanks to this, its directional derivative through $(k=k_{\rm F}, \omega=0)$ gives zero contribution in arbitrary direction and the whole $\Sigma^{G_{0} W_{0}}(k, \omega)$ is differentiable at the point $(k=k_{\rm F}, \omega=0)$. Importantly, when we take the $k$-derivative of the quasiparticle spectrum $E(k)$ in Eqs.~(\ref{eq:E-G0W0}) and (\ref{eq:E-RPA}) the derivative of the self energy is always taken along the line $\omega=b(k-k_{\rm F})$ and therefore the non-analytic term gives no contribution.

With the above formula, using Eqs.~(\ref{eq:sigma-Z}) and (\ref{eq:sigma-chi}), $\chi$ and $Z$ also have the singular contribution of the same type. Yet, their derivatives are finite at the point $(k=k_{\rm F}, \omega=0)$ and are exclusively related to $\partial \Sigma/\partial k$ and $\partial \Sigma/\partial \omega$ as
\begin{eqnarray}
\left.
\frac{\partial {\rm Re}\Sigma(k, 0)}{\partial k}
\right|_{k=k_{\rm F}}
&=&
\left.
\frac{\partial {\rm Re}\chi(k, 0)}{\partial k}
\right|_{k=k_{\rm F}}
\equiv \Delta_{\chi}
,
\\
\left.
\frac{\partial {\rm Re}\Sigma(k_{\rm F}, \omega)}{\partial \omega}
\right|_{\omega=0}
&=&
1-{\rm Re}Z(k_{\rm F}, 0)
\equiv
\Delta_{Z}
.
\end{eqnarray}
Taking the derivatives of Eqs.~(\ref{eq:E-G0W0}) and (\ref{eq:E-RPA}) and applying these formulas, we get
\begin{eqnarray}
\Delta v_{\rm F}^{{\rm G}_{0}{\rm W}_{0} }
&\equiv&
v_{\rm F}^{{\rm G}_{0}{\rm W}_{0} }
-
v_{\rm F}^{(0)}
=
\frac{\Delta_{\chi}
+k_{\rm F}\Delta_{Z}}
{1-\Delta_{Z}}
+
\delta v
\label{eq:v-G0W0}
\\
\Delta v_{\rm F}^{\rm RPA}
&\equiv&
v_{\rm F}^{\rm RPA}
-
v_{\rm F}^{(0)}
=
\Delta_{\chi}
+k_{\rm F}\Delta_{Z}
\label{eq:v-RPA}
\end{eqnarray}
with 
\begin{eqnarray}
\delta v=
\frac{{\rm Re}\Sigma(k_{\rm F}, E_0(k_{\rm F}))}{[1-\Delta_{Z}]^2}
\left.
\frac{\partial }{\partial k}
\left[
\left.
\frac{\partial {\rm Re}\Sigma}{\partial \omega}
\right|_{\omega=0}
\right]
\right|_{k=k_{\rm F}}
.
\end{eqnarray}
The order of the derivatives above is not changeable. $v_{\rm F}^{(0)}$ ($=k_{\rm F}$) is the non-interacting Fermi velocity. Notably in the RPA expression the contributions from the $\chi$ and $Z$ terms are separable.

Now our interest goes to the impacts of $\Delta_{\chi}$ and $\Delta_{Z}$. Before proceeding to numerics, let us recall some analytic results to derive an important feature of Eq.~(\ref{eq:v-RPA}). The asymptotic formula of the total $v^{\rm RPA}_{\rm F}$, which becomes exact in the dense ($r_{\rm s}\rightarrow \infty$) limit, can be derived from the formula for $\Sigma^{\rm G_{0}W_{0}}$ by Quinn and Ferrell~\cite{Quinn1958} as
\begin{eqnarray}
v^{\rm RPA}_{\rm F}
\simeq
v^{(0)}_{\rm F}
-\frac{1}{2\pi}({\rm ln}r_{\rm s}+2-{\rm log}\frac{\pi}{\alpha})
.
\label{eq:QF-formula}
\end{eqnarray}
Remarkably, in the asymptotically exact limit $v^{\rm RPA}_{\rm F}$ is larger than $v_{\rm F}^{(0)}$. We also refer to the inverse effective mass formula for later convenience
\begin{eqnarray}
\frac{1}{m^{\ast {\rm RPA}}}
=
\frac{v_{\rm F}^{\rm RPA}}{v_{\rm F}^{(0)}}
\simeq
1-\frac{\alpha r_{\rm s}}{2\pi}({\rm ln}r_{\rm s}+2-{\rm log}\frac{\pi}{\alpha})
\end{eqnarray}
Another fact is that $\Delta_{Z}$ ($=\left.\partial {\rm Re}\Sigma(k_{\rm F}, \omega)/\partial \omega\right|_{\omega=0}$) is negative definite for physically reasonable self energy~\cite{Hedin-Lundqvist}, which is a consequence of the general spectral representation of the self energy
\begin{eqnarray}
{\rm Re}\Sigma(k, \omega)
=
\Sigma_{\rm x}(k)
+
\frac{1}{\pi}\mathcal{P}
\int d\omega' \frac{|{\rm Im}\Sigma(k,\omega')|}{\omega-\omega'}
\end{eqnarray}
and holds regardless of the approximation used. $\mathcal{P}$ denotes the principal value integral. $\Sigma_{\rm x}(k)$ denotes the bare Coulomb exchange term. This is also consistent with the physical condition that the quasiparticle spectral weight $1/Z(k_{\rm F},0)$ has to be not larger than unity. Hence, in the dense homogeneous electron gas, the Coulomb interaction effect totally decrease the effective mass of electrons and, if we neglect $\Delta_{\chi}$, the mass is incorrectly increased.

\begin{figure}[h!]
 \begin{center}
  \includegraphics[scale=0.7]{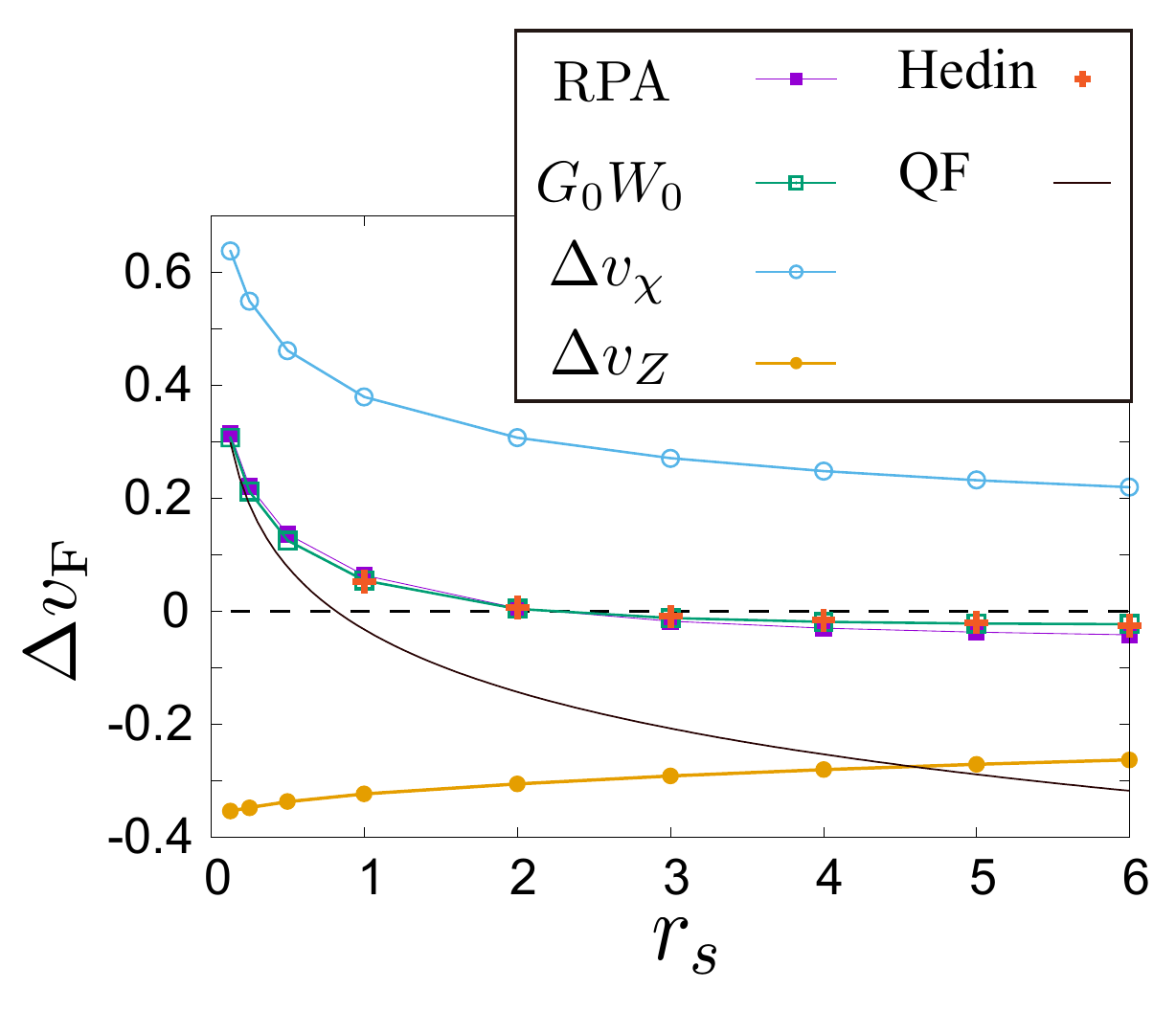}
  \caption{Corrections to the Fermi velocity. (filled and open squares) The total correction with the RPA and $G_{0}W_{0}$ formula. (open and filled circles) The corrections $\Delta v_{\chi}=\Delta_{\chi}$ and $\Delta v_{Z}=k_{\rm F}\Delta_{Z}$, respectively. (cross) Data from Ref.~\onlinecite{Hedin1965}. (line) The Quinn-Ferrell formula Eq.~(\ref{eq:QF-formula}).} 
  \label{fig:Fermi-veloc}
 \end{center}
\end{figure}

\begin{figure}[h!]
 \begin{center}
  \includegraphics[scale=0.7]{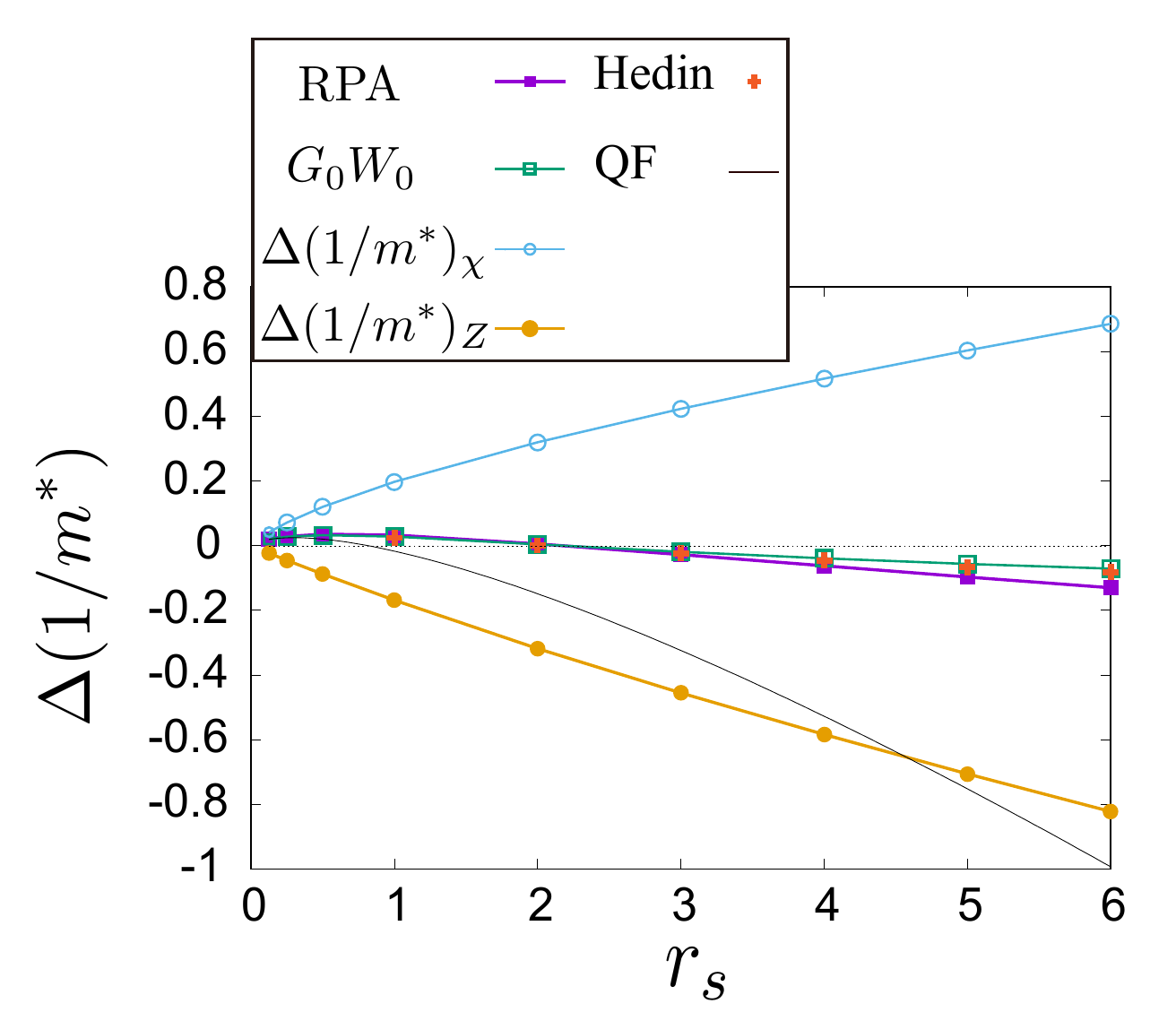}
  \caption{Corrections to the inverse effective mass. See Fig.~\ref{fig:Fermi-veloc} for legends.} 
  \label{fig:inv-mass}
 \end{center}
\end{figure}

We calculated the Fermi velocity by numerical derivatives of Eqs.~(\ref{eq:E-G0W0}) and (\ref{eq:E-RPA}) in the realistic metallic regime $r_{\rm s}\in (0, 6]$. The $G_{0}W_{0}$ self energy [Eq.~(\ref{eq:sigma-G0W0})] was computed using the recipe of Hedin~\cite{Hedin1965} on grid points in the $(k, \omega)$ space and the derivatives were evaluated with the finite-difference method. We show the results in Fig.\ref{fig:Fermi-veloc}. The total values of the Fermi velocity $v^{\rm RPA}_{\rm F}$ and $v^{\rm G_{0}W_{0}}_{\rm F}$ depart from the Quinn-Ferrell formula as $r_{\rm s}$ increases. They are wholly close in the current $r_{\rm s}$ range, allowing us to interpret the effects of $\Delta_{\chi}$ and $\Delta_{Z}$ with the separable RPA formula. In this view, we find that the large cancellation between $\Delta_{\chi}$ and $k_{\rm F}\Delta_{Z}$ occurs in the entire range. The absolute values of them are about more than five times larger than the total one in the dilute regime $r_{\rm s}>2$. In the dense regime $0.125 \lesssim r_{\rm s} \lesssim 2$ the $v_{\rm F}$ correction becomes positive as expected from Eq.~(\ref{eq:QF-formula}). The $\Delta_{\chi}$ term then becomes dominant, but $k_{\rm F}\Delta_{Z}$ still remains comparable. 

Finally we evaluated the corrections to the inverse effective mass,  $\Delta(1/m^{\ast})$ (Fig.~\ref{fig:inv-mass}). The same behavior is observed as well; small total correction ($\Delta(1/m^{\ast})^{\rm RPA}$ and $\Delta(1/m^{\ast})^{\rm G_{0}W_{0}}$) and far larger and cancelling contributions of $\Delta_{\chi}$ and $\Delta_{Z}$  ($\Delta(1/m^{\ast})^{\chi}$ and $\Delta(1/m^{\ast})^{Z}$). Remarkably, the absolute values of the latter are $\gtrsim$ 0.1, which can yield appreciable change of $m^{\ast}$ if individually considered. Their total sum does not, in agreement with the reports before~\cite{Hedin1965,Hedin-Lundqvist, Grimvall,Giuliani-Vignale,Simion-Giuliani2008}. 

We conclude that, in the RPA, the total quasiparticle Fermi velocity is a consequence of the combinatorial effects of the spatial (${\bm k}$) and temporal ($\omega$) correlations. Ignoring either of them, which respectively correspond to $\chi$ and $Z$ in the Eliashberg formalism, may yield severe error on the low-energy quasiparticle spectrum. Apart from the problem of accuracy of the RPA itself, biased treatments of $\chi$ and $Z$ does not describe the appropriate RPA physics.

\subsection{One-shot Eliashberg equations}
\label{sec:single-Eliashberg}
We next analyze how the $Z$ and $\chi$ terms affect the superconducting transition temperatures derived with the Eliashberg equations, based on the homogeneous electron gas model. The transition temperature, where $\phi$ sets in, is calculated by the solution of Eq.~(\ref{eq:Eliashberg-phi}) with $\phi$ in the denominator being zero. Namely, it can be done by solving
\begin{eqnarray}
Z_{i}([Z, \chi]; i\omega_{j})
&=&1+Z_{i}^{\rm ph}([Z, \chi]; i\omega_{j})+Z_{i}^{\rm el}([Z, \chi]; i\omega_{j})
\nonumber \\
\label{eq:Z-normal}
\\
\chi_{i}([Z, \chi]; i\omega_{j})
&=&\chi_{i}^{\rm ph}([Z, \chi]; i\omega_{j})+\chi_{i}^{\rm el}([Z, \chi]; i\omega_{j})
\label{eq:chi-normal}
\end{eqnarray}
self-consistently and next solving
\begin{eqnarray}
\phi_{i}([Z, \chi, \phi]; i\omega_{j})
&=&\phi_{i}^{\rm ph}([Z, \chi, \phi]; i\omega_{j})+\phi_{i}^{\rm el}([Z, \chi, \phi]; i\omega_{j})
\nonumber \\
\label{eq:gap-eq}
\end{eqnarray}
once. Here we have expressed the contributions related to the phonon-mediated and screened Coulomb interactions $W^{\rm ph}$ and $W^{\rm el}$ with superscripts ``ph" and ``el", respectively. 

\begin{figure*}[t!]
 \begin{center}
  \includegraphics[scale=0.5]{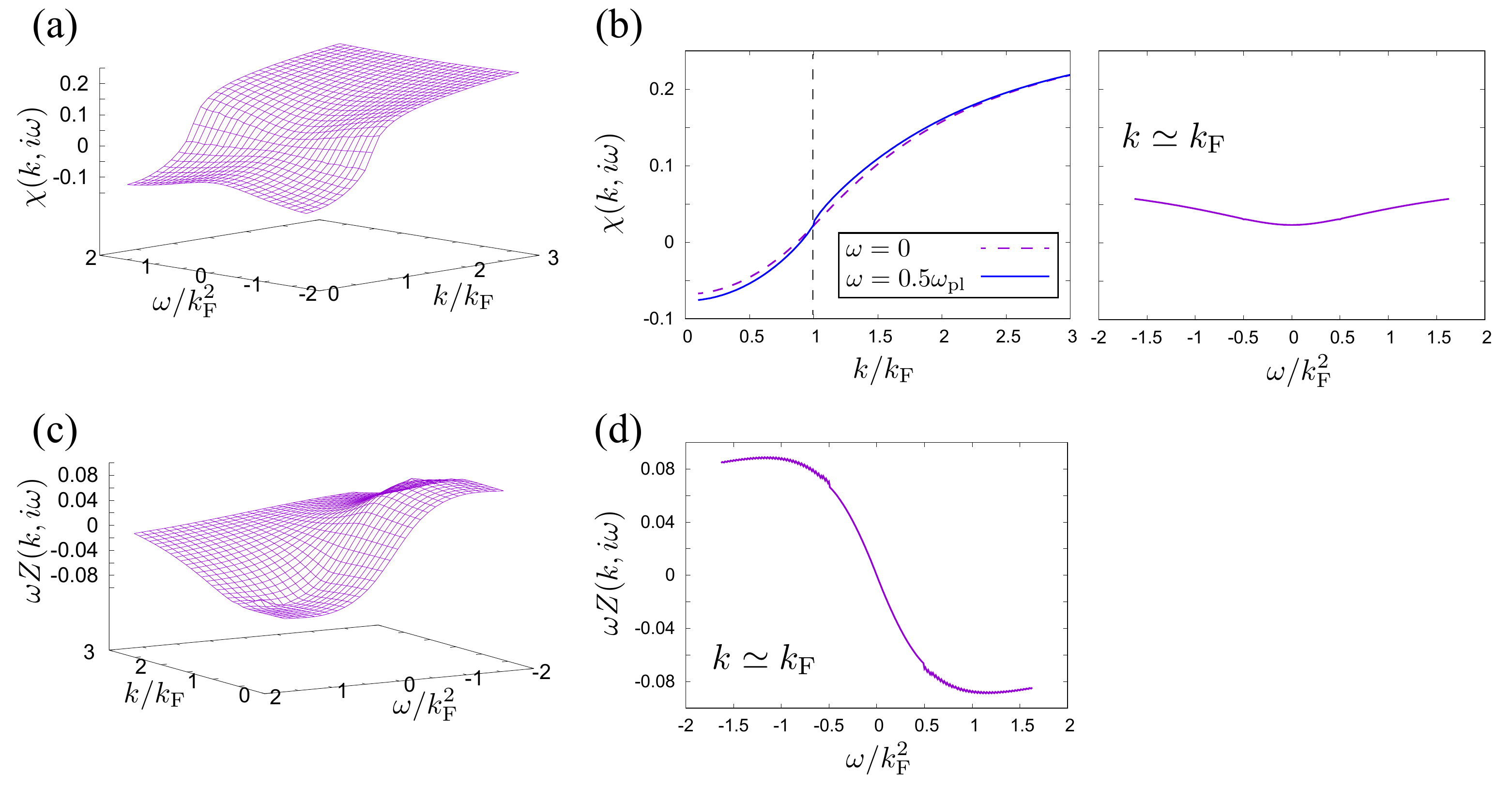}
  \caption{Calculated normal state self energy for $r_{\rm s}=3.0$. (a) Overview of $\chi(k, i\omega)$ and (b) that along specific lines. (c) Overview of $\omega Z(k, i\omega)$ and (d) that along a specific line. Subtle oscillation in (d) is due to a numerical instability of the formula at $k\simeq k_{\rm F}$, $|\omega|>\omega_{\rm thr}$ (see Appendix~\ref{sec:Hedin-imaginary} for definition of $\omega_{\rm thr}$).} 
  \label{fig:chi-Z-sum}
 \end{center}
\end{figure*}

What we would like to clarify is if the terms $Z^{\rm el}$ and $\chi^{\rm el}$, which typically have the structures in the energy scale wider than the phononic one, may change the $T_{\rm c}$ significantly. To this end, we adopt the RPA screened Coulomb interaction for $W^{\rm el}$ as in the previous subsection and the Einstein model for $W^{\rm ph}=\sum_{\nu}|g^{\nu}_{k, k'}|^2 D_{\nu}(k-k',i(\omega_{j}-\omega_{j'}))$, that is,
\begin{eqnarray}
|g^{\nu}_{k, k'}|^2
=\frac{\lambda \omega_{\rm E}}{2N_{\rm F}}
\end{eqnarray}
and
\begin{eqnarray}
D_{\nu}(k-k', i(\omega_{j}-\omega_{j'}))
=
-\frac{2\omega_{\rm E}}{(\omega_{j}-\omega_{j'})^2 + \omega_{\rm E}^2}
.
\end{eqnarray}
The complete self-consistent solution of Eqs.~(\ref{eq:Z-normal}) and (\ref{eq:chi-normal}) is still a tedious task as we explained before. Even a component $\chi^{\rm ph}$ artificially diverges with the above Einstein model approximation [see Eq.~(\ref{eq:Eliashberg-chi})]. To concentrate on our goal and keep us consistently within the non-self consistent $G_{0}W_{0}$ theory level, we approximate the original equations as follows. We first calculate $Z$ and $\chi$ non-self-consistently by
\begin{eqnarray}
Z_{i}([Z, \chi]; i\omega_{j})
&\simeq&1+Z_{i}^{\rm el}([1, 0]; i\omega_{j})
\nonumber \\
\\
\chi_{i}([Z, \chi]; i\omega_{j})
&\simeq&\chi_{i}^{\rm el}([1, 0]; i\omega_{j})
\end{eqnarray}
and later solve the equation for $\phi$  [Eq.~(\ref{eq:gap-eq})] with the energy averaging~\cite{Sanna2018}
\begin{eqnarray}
\phi(\xi, i\omega_{j})
&=&
\frac{1}{N(\xi)}\sum_{i}\delta(\xi-\xi_{i})\phi_{i}(i\omega_{j}),
\\
W(\xi, \xi', i\nu)
&=&
\frac{1}{N(\xi)N(\xi')}\sum_{ii'}\delta(\xi-\xi_{i})\delta(\xi-\xi_{i'})W_{ii'}(i\nu).
\nonumber \\
\end{eqnarray}
Equation (\ref{eq:gap-eq}) is transformed to the linear system
\begin{eqnarray}
&&\phi(\xi, i\omega_{j})
= -T\sum_{j'}
\int d\xi' N(\xi')W(\xi, \xi', i(\omega_{j}-\omega_{j'}))
\nonumber \\
&& \times
\frac{\phi(\xi', i\omega_{j'})}{\left[\omega_{j'}Z(\xi', i\omega_{j'})\right]^2 + \left[\xi'+\chi(\xi',i\omega')\right]^2}
\nonumber \\
&&
\simeq-T\sum_{j'}
\sum_{\xi'} \Delta\xi' N(\xi')W(\xi, \xi', i(\omega_{j}-\omega_{j'}))
\nonumber \\
&& \times
\frac{\phi(\xi', i\omega_{j'})}{\left[\omega_{j'}Z(\xi', i\omega_{j'})\right]^2 + \left[\xi'+\chi(\xi',i\omega')\right]^2}
\nonumber \\
&&\equiv \sum_{j'\xi'}K_{jj'}(\xi, \xi')\phi(\xi', i\omega_{j'})
.
\label{eq:Eliashberg-mat}
\end{eqnarray}
The chemical potential is tuned $\mu_{0}\rightarrow \mu_{0}+\chi(0,0)$ so that the radius of the Fermi surface is kept constant. In principle, the transition temperature $T_{\rm c}$ is given by the temperature at which the largest eigenvalue of the matrix $K_{jj'}(\xi, \xi')$, $\Lambda$, amounts to unity. To relate the superconducting instability and the mass corrections more transparently, we instead execute the following procedure: set the system parameters (summarized below) so that $\Lambda \sim O(1)$ and evaluate $\Lambda$ with and without $Z$ and $\chi$ to see how $\Lambda$ changes.


\begin{figure}[t!]
 \begin{center}
  \includegraphics[scale=0.65]{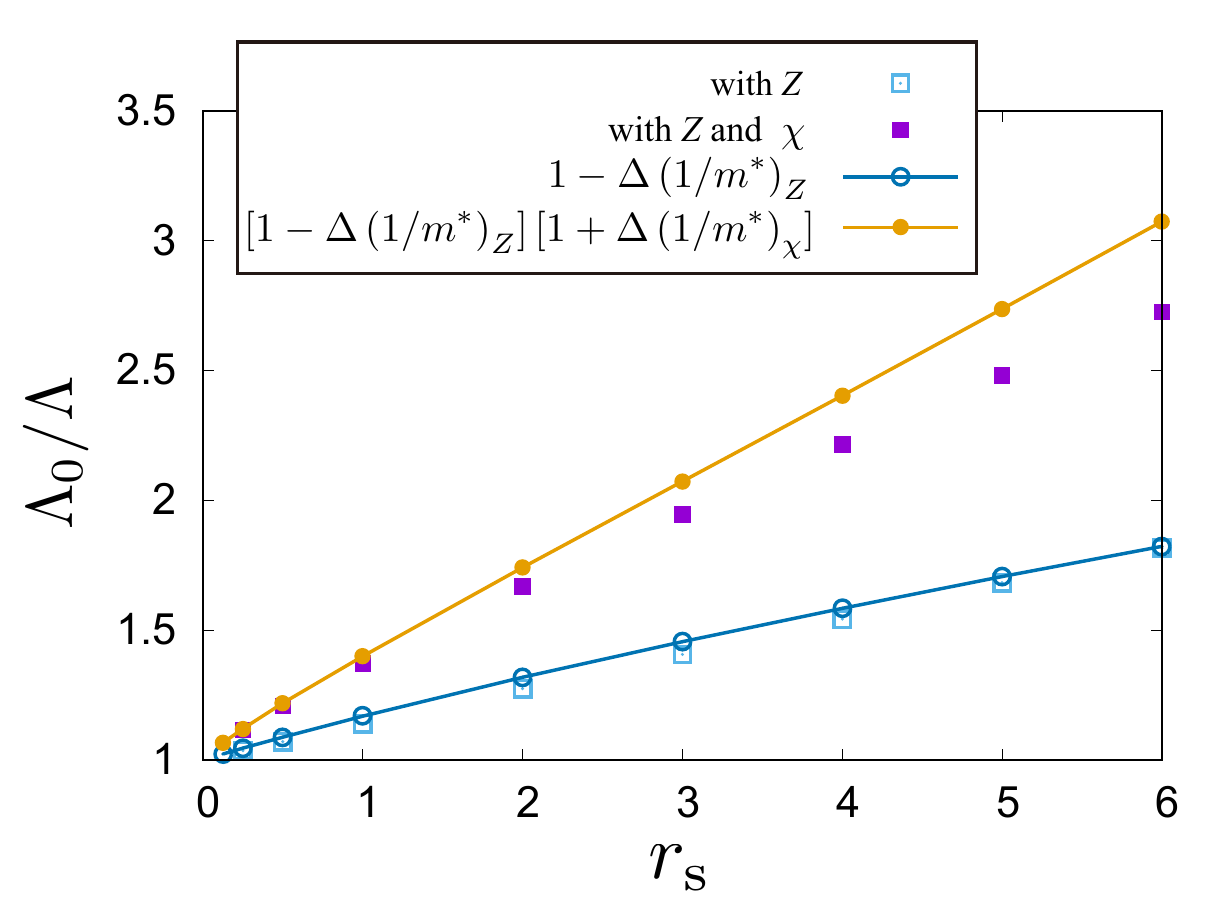}
  \caption{Ratio of the largest eigenvalues of the linear kernel in Eq.~(\ref{eq:Eliashberg-mat}) with and without the self-energy corrections. Approximate formulas are plotted with lines and points.} 
  \label{fig:eigen-ratio}
 \end{center}
\end{figure}

The current system is controlled by multiple paramaters: the coupling constant $\lambda$, Einstein frequency $\omega_{\rm E}$, plasma frequency $\omega_{\rm p}=\sqrt{3/r_{\rm s}^3}$, Fermi energy $E_{\rm F}=1/(2\alpha r_{\rm s})^2$, and temperature $T$. In this work we solve the Eliashberg equations for a specific set of parameters; $\lambda=0.5$, $\omega_{\rm E}=0.01\omega_{\rm p}$ and $T=0.05\omega_{\rm E}$ and with variable $r_{\rm s}$. This setting is realistic as it corresponds to $\omega_{\rm E}\simeq$0.167~eV and $T\simeq$97~K for $r_{\rm s}=2.0$, which may realize in compressed hydride superconductors~\cite{Flores-Livas2020}.

We calculated $\chi(k,i\omega)$ and $Z(k,i\omega)$ on the imaginary frequency axis and evaluated $\Lambda$. For the former calculation we formulated the Hedin method~\cite{Hedin1965} modified to the imaginary frequencies with the zero temperature approximation (see Appendix ~\ref{sec:Hedin-imaginary}). For the latter, nonuniform grid points for $\xi$ were generated for efficient numerical convergence. The Matsubara frequency cutoff and higher $\xi$ cutoff for the latter were set to $1.0\omega_{\rm p}$ and $1.5E_{\rm F}$, respectively. The calculated $\chi(k,i\omega)$ and $Z(k,i\omega)$ for a representative $r_{\rm s}=3.0$ are shown in Fig.~\ref{fig:chi-Z-sum}. The dependences are entirely smooth except for the derivative singularity and their slopes at the Fermi level $(k=k_{\rm F}, i\omega=0)$ are consistent with the results in the previous subsection. The derivative singularity along $k$ as described by analytic continuation of Eq.~(\ref{eq:sigma-expand}) was observed in $\chi(k,i\omega)$ [Fig.~\ref{fig:chi-Z-sum}(b)] but confirmed to be integrable.


With the knowledge of the $W^{\rm ph}$ effect in the McMillan equation Eq.~(\ref{eq:MAD}), one would expect that the ratio $\Lambda_{0}/\Lambda$, where $\Lambda_{0}$ denotes the value without the self energy effect, well correlates with $m^{\ast}$ and therefore $\Lambda_{0}/\Lambda \simeq 1$ from the results in the previous subsection. Interestingly, we observed that $Z$ and $\chi$ have non-cancelling contributions to $\Lambda_{0}/\Lambda$, both of which reduce $\Lambda$. In Fig.~\ref{fig:eigen-ratio} we show the $r_{\rm s}$ dependence of the ratio.

To track the origin of the significant reduction of $\Lambda$ we rederive the Fermi surface approximation~\cite{Scalapino1966, Allen-harmonics1976, Margine-Giustino2013,Oganov-book-Akashi}. In this approximation, we insert the identity $1=\int d\xi' \delta(\xi'-\xi_{i'})$ into the $i'$-sum in Eqs.~(\ref{eq:Eliashberg-phi}). The right hand side is then symbolically expressed as
\begin{eqnarray}
\sum_{i'}\int d\xi'\delta(\xi'-\xi_{i'})F(\xi_{i'}, i')
.
\end{eqnarray} 
The integrand is abbreviated as $F(\xi_{i'}, i')$, whose first and second arguments represent the dependences on the basis states through $\xi_{i'}$ and otherwise, respectively. The following approximation is adopted
\begin{eqnarray}
\sum_{i'}\int d\xi'\delta(\xi'-\xi_{i'})F(\xi_{i'}, i')
\simeq
\sum_{i'}\int d\xi'\delta(\xi_{i'})F(\xi', i'),
\nonumber \\
\end{eqnarray}
which is acceptable when $F(\xi_{i'}, i')$ (i) has large value only within a tiny energy range where the electronic DOS is almost constant and (ii) quickly diminishes to zero for state $i$ with large $|\xi_{i'}|$. As a result we get
\begin{eqnarray}
&&\phi_{i}(i\omega_{j})
\simeq
-T\sum_{i'j'}\int d\xi'\delta(\xi_{i'})
W_{ii'}(i(\omega_{j}-\omega_{j'}))
\nonumber \\
&&\times\frac{\phi_{i'}(i\omega_{j'})}{\left[\omega_{j'}Z_{i'}(i\omega_{j'})\right]^2+\xi'^2\left[1+\frac{\Delta_{\chi}}{v^{(0)}_{\rm F}}\right]^2+\left[\phi_{i'}(i\omega_{j'})\right]^2}
.
\end{eqnarray}
Here we have applied the linear expansion $\chi_{i'}(i\omega_{j'})\simeq \xi_{i'} \frac{\Delta_{\chi}}{v^{(0)}_{\rm F}}$. Taking the integral with respect to $\xi'$ and introducing the transformation $\phi_{i}(i\omega_{j})=Z_{i}(i\omega_{j})\Delta_{i}(i\omega_{j})$ in parallel to the standard derivation, we get
\begin{eqnarray}
&&Z_{i}(i\omega_{j})\Delta_{i}(i\omega_{j})
\nonumber \\
&&=-\frac{\pi T}{1+\frac{\Delta_{\chi}}{v^{(0)}_{\rm F}}}
\sum_{i'j'}\delta(\xi_{i'})\frac{W_{ii'}(i(\omega_{j}-\omega_{j'}))}{\omega_{j'}^2+\left[\Delta_{i'}(i\omega_{j'})\right]^2}\Delta_{i'}(i(\omega_{j'}))
\label{eq:Eliashberg-Fermi-linear}
\end{eqnarray}
Applying the linear expansion also to $\omega_{j}Z_{i}(i\omega_{j})\simeq \omega_{j}(1-\Delta_{Z})$ and comparing the prefactor of the right hand side of Eq.~(\ref{eq:Eliashberg-Fermi-linear}) with and without the self energy effects, we finally obtain
\begin{eqnarray}
\frac{\Lambda_{0}}{\Lambda}
\simeq(1-\Delta_{Z})\left(1+\frac{\Delta_{\chi}}{v^{(0)}_{\rm F}}\right)
.
\label{eq:ratio-approx}
\end{eqnarray}

In the homogeneous electron gas, we find that the curves in Fig.~\ref{fig:eigen-ratio} are approximately reproduced with the above formula, the error of which at large $r_{\rm s}$ regime may be attributed to a breakdown of the linear approximation to $\chi$ in the high-$\xi'$ regime. This formula shows that $\Delta_{Z}$ and $\Delta_{\chi}$, having the opposite signs, cooperatively reduce $\Lambda$. This behavior is in stark contrast with that of the effective mass, on which $\Delta_{Z}$ and $\Delta_{\chi}$ cancel with each other [Eq.~(\ref{eq:v-RPA})]. Careful examination of the derivation of Eq.~(\ref{eq:ratio-approx}) brings us understanding as to why $\Lambda_{0}/\Lambda$ and $m^{\ast}$ behaves differently. Namely, what actually affects the former is not the mass renormalization but total number of quasiparticle states near the Fermi level. The $Z$ term, representing the $\omega$ dependence of the self energy, simultaneously induces the mass enhancement and spectral weight reduction, with prefactors $1-\Delta_{Z}$ and $1/(1-\Delta_{Z})$, respectively. The $\chi$ term, on the other hand, reduces the mass by widening the metallic band, the latter of which results in the reduction of the normal density of state at the Fermi level $N(0)\rightarrow N(0)/(1+\frac{\Delta_{\chi}}{v^{(0)}_{\rm F}}) $; that is why the factor appears in Eq.~(\ref{eq:Eliashberg-Fermi-linear}). Since the Eliashberg gap equation Eq.~(\ref{eq:Eliashberg-mat}) has a form where the integrands, being approximately Lorentzian in $\xi'$ and $\omega_{j'}$, are integrated within the interaction energy scales, the reduction of the number of quasiparticle states results in the reduction of the total integral. Note that in the phononic theory, where $\chi^{\rm ph}$ is practically approximated to zero, interpretations based on the effective mass or number of states do not make any difference. We here also correctly state how the mass enhancement is related to the pairing strength for the phononic self energy case: the factors that renormalizes the effective mass and the spectral weight, often called {\it mass enhancement factor}, are the same, the latter of which affects the pairing strength. In some preceding careful discussions this assertion seems to have been acknowledged implicitly (e.g., Ref.~\onlinecite{Marsiglio-review}). Yet to the author's knowledge there has been no context in the phonon-mediated pairing theory where this was highlighted.

\section{Conclusions and Perspectives}
We have analyzed the self energy in the homogeneous electron gas in terms of the components of the Eliashberg theory. We have examined the effects of the $Z$ and $\chi$ terms on the effective mass and superconducting pairing instability within the non-self consistent G$_{0}$W$_{0}$ formalism. They have been usually neglected in the calculation of the Eliashberg equations for phonon-mediated superconductors. We, nevertheless, have demonstrated that they both have significant impacts. For the effective mass these effects largely cancel with each other, which corresponds to an established fact in the GW theory literature known as cancellation of the $k$-derivative and $\omega$-derivative~\cite{Hedin-Lundqvist,Grimvall,Simion-Giuliani2008, Tomczak2014}. Neglecting one of them results in artificially increased or reduced effective mass. For the superconducting instability, they cooperatively suppress the pairing via the reduction of the spectral density of states contributing to the pairing. The contrasted dependences are represented by the formulas Eqs.(\ref{eq:v-RPA}) and (\ref{eq:ratio-approx}). The current results, derived for the HEG system, indicate that both terms have to be treated on equal footing for the accurate first-principles Eliashberg calculations for a wide range of metallic systems. 

Here we have to recall that the original Eliashberg equations involve iterative update of ${\bm G}$ with $W^{\rm ph}$ included, which were neglected in the current analyses. Although the accuracy of the self-consistent calculation with the GW approximation has been a matter of debate~\cite{vonBarth-Holm1996,Shirley-GW,Aryasetiawan-review, Marom-Rubio-GW,Takada-GISC,Kutepov-GWG}, detailed study on these effect would be required in the future to establish the basis for accurate first-principles Eliashberg calculation. In particular the $W^{\rm ph}$ effect on $Z^{\rm el}$ and $\chi^{\rm el}$ would be interesting for systems where the electronic and phononic energy scales compete, such as in light-element superconductors. Low dimensional systems, where the plasmon modes can be anomalously soft, would also be an intriguing arena for such electron-phonon self energy effects.

In the Eliashberg literature, $W^{\rm el}$ in the $\chi$ and $Z$ formulas are mostly ignored as its effects are presumably included in $G_{\rm 0}$. Our results give us a quantitative insight into the drawback of this treatment. We have seen that this treatment is correct for the mass renormalization but not for the pairing renormalization. The physical quantities respecting the former would be relatively accurate with the complete neglect of the two terms; it applies to typical Eliashberg calculations where (semi)local Kohn-Sham exchange-correlation potential $V_{\rm xc}=V_{\rm xc}({\bm r})=\delta E_{\rm xc}/\delta n({\bm r})$ is used as $V_{\rm eff}$ for generating the basis states. Note that this treatment is not, however, correctly normed to the HEG limit because the local form trivially amounts to constant there and cannot incorporate the quasiparticle corrections. Use of the quasiparticle self-consistent GW method~\cite{scQPGW} for the basis would cure this. More notably, it is revealed that ignoring $\chi^{\rm el}$ and $Z^{\rm el}$ yields incorrectly strong superconducting instability, which would be finally weakened by $\chi^{\rm el}$ and $Z^{\rm el}$. This aspect is hidden in the standard procedure in which the total $W^{\rm el}$ effect is represented by the tunable parameter $\mu^{\ast}$, but becomes appreciable when one quantifies the self-energy effect from the first principles. A question then arises if the plasmon effect enhancing $T_{\rm c}$ in cooperation with phonons~\cite{Akashi2013PRL,Akashi2014JPSJ}, which is from the $\phi^{\rm el}$ term, may be largely cancelled once $Z^{\rm el}$ and $\chi^{\rm el}$ are considered. Davydov and coworkers~\cite{Davydov2020} have executed the Eliashberg equation with $Z^{\rm el}$ included, where the resulting values of $T_{\rm c}$ were generally larger than the experimental values. Including the $\chi^{\rm el}$ effect on top of it may give us better agreement. The total amount of the plasmonic pairing effect with all the self energy terms included is to be scrutinized in later studies, especially in various elemental and binary superconductors studied previously~\cite{Akashi2013PRL,Akashi2014JPSJ,Davydov2020}.

As a side issue, we have precisely discussed the non-analytic feature of the RPA ($G_0 W_0$) self energy which has not been prominently featured before. This is obviously inherited to $\chi$ and $Z$, but its effects are yet to be clarified. At least this does not affect the quasiparticle since the non-analytic component is smooth along $\omega=E(k)$ as discussed above. The divergence of the Coulomb potential at ${\bm k}=0$, which is the origin of the singularity, is often managed with the truncation of the Coulomb potential~\cite{Specner-Alavi-truncateCoulomb,Nakamura-GWc2016}; one then no longer encounter the present singularity. This treatment is justfied by a reasonable assumption that this singularity, being integrable, does not have any effects in the physical quantities defined by integrating the self energy. An example is the Eliashberg pairing strength as confirmed in this study. Care must be taken when one treat the self energy itself, or particularly, its formal derivatives.



\section*{Acknowledgment}
The author thanks to Ryotaro Arita, Takuya Nomoto, Youhei Yamaji, and Yusuke Nomura for fruitful discussions. This work was supported by JSPS KAKENHI Grant Numbers 20K20895 from Japan Society for the Promotion of Science (JSPS). 

\appendix
\section{Derivation of the singular term in Eq.~(\ref{eq:sigma-expand})}
\label{app:singular}
Here we extract the singular terms of the $G^{0}W^{0}$ self energy. First, the bare Coulomb exchange part is given by
\begin{eqnarray}
\Sigma_{\rm x}({\bm k}, \omega)
&=&
i
\int\frac{d^3 k'}{(2\pi)^3}
\int\frac{d\omega'}{(2\pi)}\frac{e^{i\eta(\omega + \omega')}}{\omega + \omega' -E_{0}({\bm k}+{\bm k}')(1-i\eta)}
\frac{4\pi}{k'^2}
\nonumber
\\
&=&
-
\int\frac{d^3 k'}{(2\pi)^3}\frac{4\pi}{k'^2}
\theta(k_{\rm F}-|{\bm k}+{\bm k}'|)
\nonumber
\\
&\equiv&
\Sigma_{\rm x}(k)
.
\end{eqnarray}
This integral is analytically performed and we have the well-known Fock form
\begin{eqnarray}
\Sigma_{\rm x}(k)
=
-\frac{1}{2\pi}
\left(
\frac{k_{\rm F}^2-k^2}{k}
{\rm log}
\left|
\frac{k_{\rm F}+k}{k_{\rm F}-k}
\right|
+2k_{\rm F}
\right)
\end{eqnarray}

The singular behavior originates from the small wavenumber regime of the integral. To illustrate this we introduce the cutoff for $k'$ as $k_{\rm c}$ and the non-analytic part is extracted as follows
\begin{eqnarray}
\Sigma^{\rm na}_{\rm x}(k; k_{\rm c})
\sim
-\frac{1}{\pi}
\int_{0}^{k_{\rm c}} dk'
\int_{-1}^{1} d\cos x
\theta(k_{\rm F}-|{\bm k}+{\bm k}'|)
.
\nonumber \\
\end{eqnarray}
$x$ is the polar angle between the vectors ${\bm p}$ and ${\bm k}$. Applying the Taylor expansion to the argument of the theta function, the integral is executed straightforwardly and we get to
\begin{eqnarray}
\Sigma^{\rm na}_{\rm x}(k; k_{\rm c})
\sim
-\frac{1}{\pi}
(k_{\rm F}-k)
\log
\left|
\frac{k_{\rm c}}{(k_{\rm F}-k)}
\right|
+O(k_{\rm F}-k)
.
\nonumber \\
\end{eqnarray}

Next we consider the screened Coulomb part. The long-wavelength limit of the Lindhard formula, 
\begin{eqnarray}
\frac{1}{\epsilon({\bm k}, \omega)}
\simeq
1+
\frac{\omega_{\rm p}^2}{\{\omega-\omega_{\rm p}+i\eta\}\{\omega+\omega_{\rm p}-i\eta\}}
,
\end{eqnarray}
is enough for analyzing the target singularity coming from the internal momentum $k'\lesssim k_{\rm c}$ with $\omega'\neq 0$. The infinitesimal pole shifts are introduced so that the positive and negative energy plasmon modes propagate forward and backward in time, respectively~\cite{Quinn1958}.

The correlation part of the $G^{0}W^{0}$ self energy reads
\begin{eqnarray}
\Sigma_{\rm c}({\bm k}, \omega)
&=&
i
\int\frac{d^3 k'}{(2\pi)^3}
\int\frac{d\omega'}{(2\pi)}\frac{1}{\omega + \omega' -E_{0}({\bm k}+{\bm k}')(1-i\eta)}
\nonumber \\
&&\hspace{20pt}
\times
\frac{4\pi }{k'^{2}}\left(\frac{1}{\epsilon({\bm k}', \omega')}-1\right)
\nonumber \\
&\simeq&
\int\frac{d^3 k'}{(2\pi)^3}
\frac{2\pi  \omega_{\rm p}}{k'^{2}}
\left[
\frac{\theta(|{\bm k}+{\bm k}'|-k_{\rm F})}{\omega - \omega_{p} -E({\bm k}+{\bm k}')}
\right.
\nonumber \\
&&
\hspace{40pt}
\left.
+
\frac{\theta(k_{\rm F}-|{\bm k}+{\bm k}'|)}{\omega + \omega_{p} -E_{0}({\bm k}+{\bm k}')}
\right]
.
\end{eqnarray} 
The singular contribution is similarly extracted by limiting the range of integral to $k'\lesssim k_{\rm c}$. We can assume $E_{0}({\bm k}+{\bm k}')\simeq E_{0}({\bm k})$ in the integrand and, after lengthy but straightforward calculations, we get to
\begin{eqnarray}
\Sigma^{\rm na}_{\rm c}({\bm k}, \omega; k_{\rm c})
&\sim&
\frac{\omega_{\rm p}}{2\pi}
\left[
-\frac{1}{\omega-\omega_{\rm p}-\frac{k^2-k_{\rm F}^{2}}{2}}
+
\frac{1}{\omega+\omega_{\rm p}-\frac{k^2-k_{\rm F}^{2}}{2}}
\right]
\nonumber \\
&&\hspace{20pt}
\times
(k_{\rm F}-k)
\log
\left|
\frac{k_{\rm c}}{k_{\rm F}-k}
\right|
\nonumber \\
&&
+O(k_{\rm F}-k, \omega)
.
\end{eqnarray}
Adding $\Sigma^{\rm na}_{\rm x}$, we finally obtain
\begin{widetext}
\begin{eqnarray}
\Sigma^{G^{0}W^{0},{\rm na}}(k,\omega; k_{\rm c})
&=&
\Sigma_{\rm x}^{\rm na}(k; k_{\rm c})
+
\Sigma_{\rm c}^{\rm na}(k,\omega; k_{\rm c})
\nonumber \\
&\sim&
\frac{1}{2\pi}
\left[
\frac{\omega_{\rm p}}{\omega_{\rm p}-(\omega-\frac{k^2-k_{\rm F}^{2}}{2})}
+
\frac{\omega_{\rm p}}{\omega_{\rm p}+(\omega-\frac{k^2-k_{\rm F}^{2}}{2})}
-2
\right]
(k_{\rm F}-k)
\log
\left|
\frac{k_{\rm c}}{k_{\rm F}-k}
\right|
+O(k_{\rm F}-k, \omega)
.
\end{eqnarray}
\end{widetext}

\section{$\Sigma^{G_{0}W_{0}}(k, i\omega_{j})$ in the homogeneous electron gas}
\label{sec:Hedin-imaginary}
The self energy in the normal-state homogeneous electron gas, $\Sigma^{G_{0}W_{0}}(k, i\omega_{j})$, is
\begin{widetext}
\begin{eqnarray}
\Sigma^{G_{0}W_{0}}(k, i\omega_{j})
=-T\sum_{j'}\int \frac{d^3 k'}{(2\pi)^3}\frac{V({\bm k}')}{\varepsilon({\bm k}', i\omega_{j'})}\frac{e^{-i\eta\omega_{j'}}}{i(\omega_{j}-\omega_{j'})-\xi_{{\bm k}-{\bm k}'}}
.
\end{eqnarray}
Practically we calculate this with a procedure similar to Hedin's in Ref.~\onlinecite{Hedin1965}. Namely, the $\omega_{j'}$-summation with weight $1/\varepsilon({\bm k}', i\omega_{j'})$ is decomposed as follows:
\begin{eqnarray}
&&T\sum_{j'}\frac{1}{\varepsilon({\bm k'}, i\omega_{j'})}(\cdots)
=
T\sum_{j'}\frac{1}{\varepsilon({\bm k'}, 0)}(\cdots)
+
T\sum_{|\omega_{j'}|\geq\omega_{\rm thr}}
\left(
1
-
\frac{1}{\varepsilon({\bm k'}, 0)}
\right)(\cdots)
+
T\sum_{|\omega_{j'}|\geq\omega_{\rm thr}}
\left(
\frac{1}{\varepsilon({\bm k'}, i\omega_{j'})}
-
1
\right)(\cdots)
\nonumber \\
&&\hspace{120pt}+
T\sum_{|\omega_{j'}|<\omega_{\rm thr}}
\left(
\frac{1}{\varepsilon({\bm k'}, i\omega_{j'})}
-
\frac{1}{\varepsilon({\bm k'}, 0)}
\right)(\cdots).
\label{eq:Hedin-decompose}
\end{eqnarray} 
The self energy is decomposed as well
\begin{eqnarray} 
\Sigma^{G_{0}W_{0}}(k, i\omega_{j})
=
\Sigma_{\rm c}(k)
+
\Sigma_{\rm rm}(k, i\omega_{j})
+
\Sigma_{\rm dr1}(k, i\omega_{j})
+
\Sigma_{\rm dr2}(k, i\omega_{j})
,
\end{eqnarray} 
where the terms correspond to the above integrals in order, respectively.

Taking the zero-temperature limit $(T\sum_{j'}\rightarrow\int\frac{d\omega'}{2\pi})$ and introducing the dimensionless variable $q=k/k_{\rm F}$ and $\nu=k/k^2_{\rm F}$, we can take some part of the integrals analytically. First we have
\begin{eqnarray}
\Sigma_{\rm c}(q)
=
-
\frac{k_{\rm F}}{2\pi^2}\int_{0}^{1}dq'\frac{1}{\varepsilon(q', 0)}\theta(1-|{\bm q}-{\bm q}'|)
-\frac{k_{\rm F}}{\pi}\int_{0}^{\infty}dq'\left(1-\frac{1}{\varepsilon(q',0)}\right)
.
\end{eqnarray}
The second term, screened Coulomb hole, originates from the contribution of order $\sim \int d\omega' \sin \eta \omega'/\omega'$ in the latter three terms in Eq.~(\ref{eq:Hedin-decompose}). We also have
$
\Sigma_{\rm rm}(k, i\omega_{j})
=
\Sigma'_{\rm rm}(k, i\omega_{j})
+i\Sigma''_{\rm rm}(k, i\omega_{j})
$
,
$
\Sigma_{\rm dr1}(k, i\omega_{j})
=
\Sigma'_{\rm dr1}(k, i\omega_{j})
+i\Sigma''_{\rm dr1}(k, i\omega_{j})
$
,
$
\Sigma_{\rm dr2}(k, i\omega_{j})
=
\Sigma'_{\rm dr2}(k, i\omega_{j})
+i\Sigma''_{\rm dr2}(k, i\omega_{j})
$
with
\begin{eqnarray}
&&
\Sigma'_{\rm rm}(q, i\nu)
=
-\frac{k_{\rm F}}{4\pi^2}
\int_{0}^{\infty} dq' 
\left(
1
-
\frac{1}{\varepsilon(q', 0)}
\right)
\frac{1}{qq'}
\nonumber \\
&&
\times
\left[
-(\nu_{\rm thr}-\nu)
{\rm log}
\left|
\frac{
(\nu_{\rm thr}-\nu)^2
+\frac{1}{4}
((q-q')^2-1)^2
}
{
(\nu_{\rm thr}-\nu)^2
+\frac{1}{4}
((q+q')^2-1)^2
}
\right|
-(\nu_{\rm thr}+\nu)
{\rm log}
\left|
\frac{
(\nu_{\rm thr}+\nu)^2
+\frac{1}{4}
((q-q')^2-1)^2
}
{
(\nu_{\rm thr}+\nu)^2
+\frac{1}{4}
((q+q')^2-1)^2
}
\right|
\right.
\nonumber \\
&&
+|(q-q')^2-1|
\left(\frac{\pi}{2}(2-{\rm sgn}(\nu_{\rm thr}-\nu)-{\rm sgn}(\nu_{\rm thr}+\nu))+\arctan\frac{\frac{1}{2}|(q-q')^2-1|}{\nu_{\rm thr}-\nu}
+\arctan\frac{\frac{1}{2}|(q-q')^2-1|}{\nu_{\rm thr}+\nu}
\right)
\nonumber \\
&&
\left.
-|(q+q')^2-1|
\left(\frac{\pi}{2}(2-{\rm sgn}(\nu_{\rm thr}-\nu)-{\rm sgn}(\nu_{\rm thr}+\nu))+\arctan\frac{\frac{1}{2}|(q+q')^2-1|}{\nu_{\rm thr}-\nu}
+\arctan\frac{\frac{1}{2}|(q+q')^2-1|}{\nu_{\rm thr}+\nu}
\right)
\right]
,
\end{eqnarray}
\begin{eqnarray}
\Sigma_{\rm rm} ''(q, \nu)
&=&
-\frac{k_{\rm F}}{2\pi^2}
\int_{0}^{\infty} dq' 
\left(
1
-
\frac{1}{\varepsilon(q', 0)}
\right)
\frac{1}{qq'}
\nonumber \\
&&
\left[
\frac{1}{4}((q-q')^2-1)
{\rm log}
\frac{\frac{1}{4}((q-q')^2-1)+(\nu-\nu_{\rm thr})^2}{\frac{1}{4}((q-q')^2-1)+(\nu+\nu_{\rm thr})^2}
\right.
\nonumber \\
&&\hspace{80pt}
\left.
+(\nu-\nu_{\rm thr})\arctan\frac{\frac{1}{2}((q-q')^2-1)}{\nu-\nu_{\rm thr}}
-(\nu+\nu_{\rm thr})\arctan\frac{\frac{1}{2}((q-q')^2-1)}{\nu+\nu_{\rm thr}}
\right.
\nonumber \\
&&
\left.
-\frac{1}{4}((q+q')^2-1){\rm log}\frac{\frac{1}{4}((q+q')^2-1)+(\nu-\nu_{\rm thr})^2}{\frac{1}{4}((q+q')^2-1)+(\nu+\nu_{\rm thr})^2}
\right.
\nonumber \\
&&\hspace{80pt}
\left.
-(\nu-\nu_{\rm thr})\arctan\frac{\frac{1}{2}((q+q')^2-1)}{\nu-\nu_{\rm thr}}
+(\nu+\nu_{\rm thr})\arctan\frac{\frac{1}{2}((q+q')^2-1)}{\nu+\nu_{\rm thr}}
\right]
.
\end{eqnarray}
\begin{eqnarray}
&&\Sigma'_{\rm dr1}(q, i\nu)
=
-\frac{k_{\rm F}}{4\pi^2}
\int_{0}^{\infty} dq' 
\frac{1}{qq'}
\int_{\nu_{\rm thr}}^{\infty}d\nu'
\left(
\frac{1}{\varepsilon(q', i\nu')}
-1
\right)
\left[
{\rm log}
\left|
\frac{
(\nu-\nu')^2
+\frac{1}{4}
((q-q')^2-1)^2
}
{
(\nu-\nu')^2
+\frac{1}{4}
((q+q')^2-1)^2
}
\right|
\right.
\nonumber \\
&&\hspace{160pt}
\left.
+
{\rm log}
\left|
\frac{
(\nu+\nu')^2
+\frac{1}{4}
((q-q')^2-1)^2
}
{
(\nu+\nu')^2
+\frac{1}{4}
((q+q')^2-1)^2
}
\right|
\right]
,
\end{eqnarray}
\begin{eqnarray}
\Sigma''_{\rm dr1}(q,i\nu)
&=&
-\frac{k_{\rm F}}{2\pi^2}
\int_{0}^{\infty} dq' 
\int_{\nu_{\rm thr}}^{\infty}d\nu'
\left(
\frac{1}{\varepsilon(q', i\nu')}
-1
\right)
\frac{1}{qq'}
\left[
\arctan\frac{\frac{1}{2}((q-q')^2-1)}{\nu-\nu'}
+
\arctan\frac{\frac{1}{2}((q-q')^2-1)}{\nu+\nu'}
\right.
\nonumber \\
&&
\hspace{160pt}
\left.
-\arctan\frac{\frac{1}{2}((q+q')^2-1)}{\nu-\nu'}
-\arctan\frac{\frac{1}{2}((q+q')^2-1)}{\nu+\nu'}
\right],
\end{eqnarray}

\begin{eqnarray}
&&\Sigma'_{\rm dr2}(q, i\nu)
=
-\frac{k_{\rm F}}{4\pi^2}
\int_{0}^{\infty} dq' 
\frac{1}{qq'}
\int_{0}^{\nu_{\rm thr}}
\left(
\frac{1}{\varepsilon(q', i\nu')}
-
\frac{1}{\varepsilon(q', 0)}
\right)
\left[
{\rm log}
\left|
\frac{
(\nu-\nu')^2
+\frac{1}{4}
((q-q')^2-1)^2
}
{
(\nu-\nu')^2
+\frac{1}{4}
((q+q')^2-1)^2
}
\right|
\right.
\nonumber \\
&&\hspace{160pt}
\left.
+
{\rm log}
\left|
\frac{
(\nu+\nu')^2
+\frac{1}{4}
((q-q')^2-1)^2
}
{
(\nu+\nu')^2
+\frac{1}{4}
((q+q')^2-1)^2
}
\right|
\right]
\end{eqnarray}

\begin{eqnarray}
\Sigma''_{\rm dr2}(q,i\nu)
&=&
-\frac{k_{\rm F}}{2\pi^2}
\int_{0}^{\infty} dq' 
\int_{0}^{\nu_{\rm thr}}d\nu'
\left(
\frac{1}{\varepsilon(q', i\nu')}
-
\frac{1}{\varepsilon(q', 0)}
\right)
\frac{1}{qq'}
\left[
\arctan\frac{\frac{1}{2}((q-q')^2-1)}{\nu-\nu'}
+
\arctan\frac{\frac{1}{2}((q-q')^2-1)}{\nu+\nu'}
\right.
\nonumber \\
&&
\hspace{160pt}
\left.
-\arctan\frac{\frac{1}{2}((q+q')^2-1)}{\nu-\nu'}
-\arctan\frac{\frac{1}{2}((q+q')^2-1)}{\nu+\nu'}
\right].
\end{eqnarray}

\end{widetext}
Terms $Z(k, i\omega)$ and $\chi(k, i\omega)$ entering the Eliashberg equations are related to those forms by
\begin{eqnarray}
\omega Z(k, i\omega)
&=&{\rm Im}\Sigma(k, i\omega)
,
\\
\chi (k, i\omega)
&=&{\rm Re}\Sigma(k, i\omega)
.
\end{eqnarray}

\bibliography{reference}

\end{document}